\documentclass[9pt,twocolumn,twoside]{osajnl}

\journal{ao} 

\setboolean{shortarticle}{false} 

\title{Designs for a large-aperture telescope \\
to map the CMB $10\times$ faster}

\author[1]{Michael D. Niemack}

\affil[1]{Cornell University Physics Department, Ithaca, NY 14850, USA; niemack@cornell.edu}

\dates{Compiled \today}

\ociscodes{(110.6770) Telescopes; (350.1260) Astronomical Optics; (350.4010) Microwaves; (040.1240) Detector Arrays.}

\doi{\url{http://dx.doi.org/XXXXXX}}

\begin{abstract}
Current large-aperture cosmic microwave background (CMB) telescopes have nearly maximized the number of detectors that can be illuminated while maintaining diffraction-limited image quality. The polarization-sensitive detector arrays being deployed in these telescopes in the next few years will have roughly $10^4$ detectors. Increasing the mapping speed of future instruments by at least an order of magnitude is important to enable precise probes of the inflationary paradigm in the first fraction of a second after the big bang and provide strong constraints on cosmological parameters. The CMB community has begun planning a next generation ``Stage IV'' CMB project that will be comprised of multiple telescopes with between $10^5$ -- $10^6$ detectors to pursue these goals. This paper introduces new crossed Dragone telescope and receiver optics designs that increase the usable diffraction-limited field-of-view, and therefore the mapping speed, by an order of magnitude compared to the upcoming generation of large-aperture instruments. Polarization systematics  and engineering considerations are presented, including a preliminary receiver model to demonstrate that these designs will enable high efficiency illumination of $>10^5$ detectors in a next generation CMB telescope.
\end{abstract}

\setboolean{displaycopyright}{false}

\begin{document}

\maketitle
\thispagestyle{fancy}
\ifthenelse{\boolean{shortarticle}}{\abscontent}{}

Recent measurements of the Cosmic Microwave Background (CMB) led to detections of several new cosmological signals, including gravitational lensing of the CMB, approximately mass-limited galaxy cluster catalogs, and, most recently, the ``B-mode'' polarization \cite[e.g.][]{hanson/etal:2013, bkp:2015, polarbear:2014, vanengelen/etal:2015}. This rapid progress has been enabled by the development of large arrays of background-limited low-temperature detectors. These detector arrays are typically designed to fill a large fraction of the diffraction-limited field-of-view (DLFOV) on each telescope. The upcoming generation of large-aperture ($>2$\,meter) CMB instruments includes: the Simons Array of three 2.5\,meter telescopes \cite{arnold/etal:2014}, Advanced ACTPol on the 6\,meter Atacama Cosmology Telescope \cite{henderson/etal:2015}, and SPT-3G on the 10\,meter South Pole Telescope \cite{benson/etal:2014}. Each will nearly fill the DLFOV of  these existing telescopes with roughly $10^4$ detectors operating between 30\,GHz -- 300\,GHz.

Beyond this coming generation of instruments, there is potential for substantial improvements in constraints on signals from inflationary gravity waves, neutrino properties, and other cosmological parameters if CMB measurements can be made with sufficient sensitivity and angular resolution \cite[e.g.,][]{abazajian/etal:2015, abazajian/etal:2015b, mueller/etal:2015}. The CMB community has begun planning for a ``Stage IV'' CMB survey with between $10^5$ -- $10^6$ detectors to achieve these scientific goals, including a precise probe of the inflationary tensor-to-scalar ratio, $r < 0.001$ \cite{abazajian/etal:2015}. Since current telescopes do not have sufficient throughput to illuminate $>10^5$ detectors, it is important to develop designs for higher-throughput telescopes. 

We explore a new parameter space of CMB telescope designs that offer roughly $10\times$ greater throughput than existing telescopes. In \S\ref{sec:CD} we quantify the diffraction-limited throughput tradeoffs between crossed-Dragone telescope designs. In \S\ref{sec:reimage} we present two designs for crossed-Dragone telescopes with close-packed reimaging optics that can achieve $\sim$$10\times$ faster mapping speed than upcoming CMB instruments with $\sim$$10^4$ detectors.\footnote{We note that the mapping speed of a CMB instrument scales linearly with the number of detectors assuming a constant noise level per detector, or as the inverse square of the instrument noise equivalent temperature, $NET^{-2}$.} \S\ref{sec:systematics} and \S\ref{sec:eng} describe systematic and engineering considerations for these designs, and we conclude in \S\ref{sec:conclusion}.

\section{Crossed-Dragone Telescopes}
\label{sec:CD}

The crossed-Dragone (CD) telescope \cite{dragone:1978} has recently been recognized as an excellent choice for CMB polarimetery \cite[for an overview, see][]{hanany/niemack/page:2013} because it provides a large DLFOV with excellent polarization fidelity \cite{mizuguch/etal:1976,tran/etal:2008}. CD telescopes were used for CMB measurements for the first time in the last decade on two small aperture ($<2$\,m) instruments \cite{bischoff/etal:2013,essinger-hileman/etal:2009}, and were proposed in a concept paper for NASA's Inflation Probe\cite{tran/etal:2010}. All of these designs have small focal ratios ($f < 2$) and place the detector arrays directly at the telescope focus. This presents a challenging baffling problem due to substantial spillover past both the primary and secondary mirrors, which has been mitigated by use of absorbing baffles. In the Atacama B-mode Search \cite{essinger-hileman/etal:2009} the telescope mirrors and baffles are cooled to $\sim$$4$\,K temperatures to minimize loading and photon noise from the baffles. In the Q/U Imaging Experiment \cite{bischoff/etal:2013} coherent detectors are used to directly measure the polarization signal, thereby reducing sensitivity to unpolarized emission from the ambient temperature baffles. 

\begin{figure}[t]
\centering
\fbox{\includegraphics[width=\linewidth]{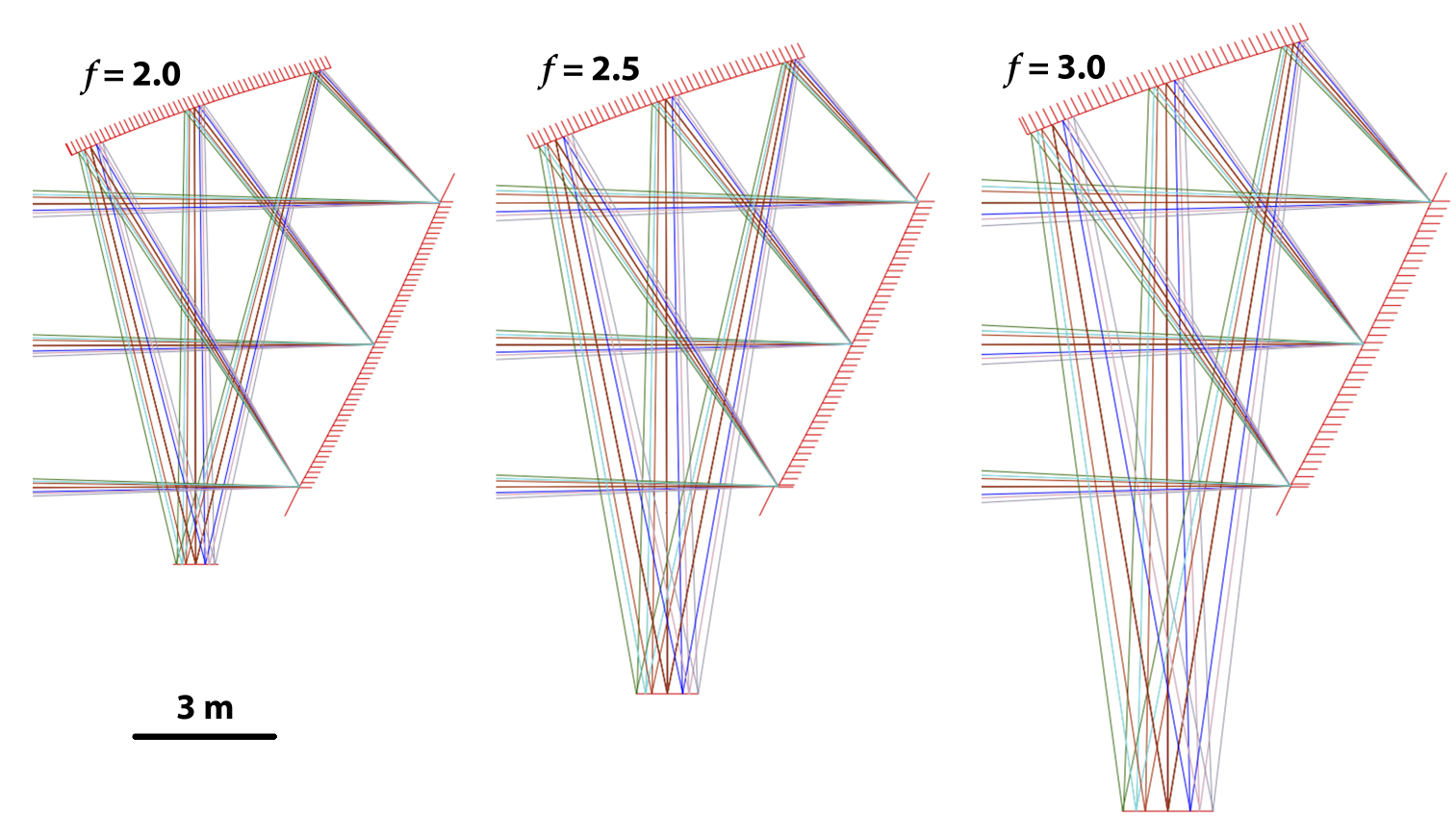}}
\caption{Three 6\,m aperture telescope designs with different $f$. Plotted rays span the 150\,GHz CFOV with Strehl ratios $>0.70$.}
\label{fig:telescopes}
\vspace{-.15in}
\end{figure}

\begin{figure}[t]
\centering
\vspace{-.05in}
\includegraphics[width=\linewidth]{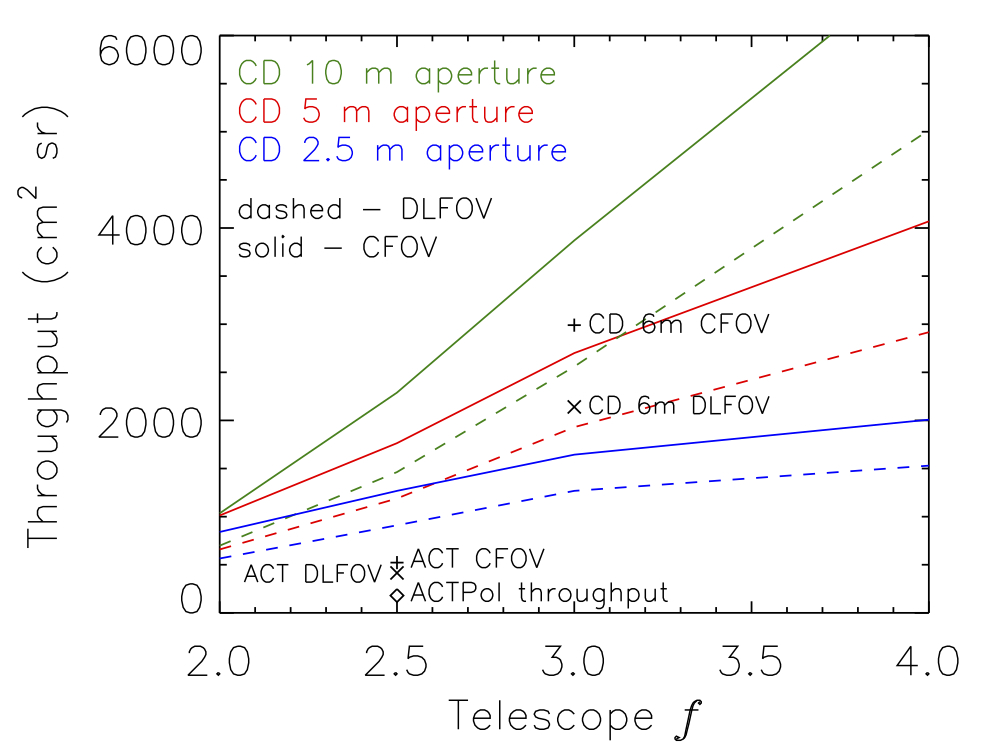}
\vspace{-.2in}
\caption{Telescope $f$ versus throughput for different telescope apertures (2.5\,m - blue/lowest, 5\,m - red, 10\,m - green/highest). Solid (dashed) lines indicate the telescope 150\,GHz CFOV (DLFOV). Also shown are points indicating the 150\,GHz throughput of the ACT FOVs, ACTPol, and the CD 6\,m design in Fig.~\ref{fig:folded} and \ref{fig:hex}.}
\label{fig:throughput}
\vspace{-.2in}
\end{figure}

Here we present CD telescope designs with larger apertures ($>2$\,m) and focal ratios ($2 < f < 3$) that mitigate the CD baffling challenges by coupling to reimaging optics with a cryogenic Lyot stop at an image of the primary mirror. Similar Lyot stops are used in both current \cite{polarbear:2014, niemack/etal:2010} and upcoming \cite{arnold/etal:2014, benson/etal:2014, henderson/etal:2015} large-aperture CMB instruments, because they can suppress spillover outside the main beam by over an order of magnitude and dramatically reduce baffling requirements. Fig.~\ref{fig:telescopes} shows three example CD designs with different $f$, and Fig.~\ref{fig:throughput} depicts the substantial increase in DLFOV as the $f$ and aperture are increased. The implementation of these designs follows \cite{granet:2001} in which the following five parameters are chosen to define the design: the primary diameter, $D_m$, the angle between the mirror and focal plane coordinate systems, $\theta_p$, the primary mirror focal length, $F$, the distance between the secondary mirror and focal plane, $L_s$, and the half angle between the focal plane and secondary mirror, $\theta_e$, which defines the focal ratio, $f = [2 \tan(\theta_e)]^{-1}$. Table~\ref{tab:params} provides the design parameters for the telescopes in Fig.~\ref{fig:telescopes} and \ref{fig:folded}. In practice, these parameters are implemented in $Code$ $V$ ray tracing software, then the distances between primary and secondary as well as secondary and focal plane are optimized to improve the image quality across a wide FOV.

The DLFOV for each telescope design is calculated by adjusting the major and minor axis of an elliptical FOV until the 150\,GHz Strehl ratios of the extreme $\pm X$ and $\pm Y$ fields are $0.80\pm 0.01$, which is commonly regarded as the minimum ``diffraction-limited'' Strehl ratio. The throughput is then calculated as the area times the solid angle at both the telescope aperture and focal plane for consistency. A similar approach is used to calculate the throughput for different wavelengths (100\,GHz, 230\,GHz, and 300\,GHz) and minimum Strehl ratios (0.70 and 0.90) shown in Fig.~\ref{fig:throughput} and \ref{fig:hex}. As described below, we find that reimaging optics can be used to correct local aberrations, and thereby improve the minimum Strehl ratios in each field from $0.70$ to $>0.80$.  We therefore define the ``correctable'' field-of-view (CFOV) as the field-of-view with Strehl ratios $>0.70$.

In Fig.~\ref{fig:throughput} we also compare the throughput of CD telescopes to the 6\,m off-axis, numerically-optimized, aplanatic Gregorian design used for the Atacama Cosmology Telescope (ACT, \cite{fowler/etal:2007}).\footnote{See  \cite{hanany/niemack/page:2013} for a comparison of the ACT and South Pole Telescope \cite{padin/etal:2008} optics.} We find that CD designs have substantially larger DLFOV and CFOV than the ACT; specifically, the 6\,m $f =3$ designs in Fig.~\ref{fig:folded} have $>5\times$ larger DLFOV than the ACT and the receiver optics presented here have $>10\times$ larger throughput than the ACTPol \cite{niemack/etal:2010} throughput shown in Fig.~\ref{fig:throughput}.

We highlight two 6\,m telescope designs that combine sufficiently large throughput to achieve $10\times$ faster mapping speed than upcoming instruments \cite{arnold/etal:2014, benson/etal:2014, henderson/etal:2015} with sufficient resolution to pursue the full range of CMB survey science, from inflation to galaxy clusters. Fig.~\ref{fig:folded} shows both $f = 3$ telescope designs in which a flat tertiary mirror has been added to fold the optics and make the designs more compact. This moves the receiver closer to the axis of rotation of the telescope when observing at a nominal elevation of 45$^\circ$. The two $f = 3$ designs in Fig.~\ref{fig:folded} have different spacings between the primary, secondary, tertiary, and receiver, which can be easily adjusted to provide space for baffling without significant changes in throughput, as indicated in Table~\ref{tab:params}. 

\begin{table}[b]
\vspace{-.3in}
\centering
\caption{Variable telescope parameters for Fig.~\ref{fig:telescopes} and \ref{fig:folded}.  Constant parameters are $D_m = 6.0$\,m and $\theta_p = 90^{\circ}$ as defined in \cite{granet:2001}. Throughput, $A\Omega$, is given for the 150\,GHz CFOV.}
\begin{tabular}{cccccc}
\hline
Figure & $f$  & $F$ (m) & $\theta_e$ ($^{\circ}$) & $L_s$ (m) & $A\Omega$ (cm$^2$ sr) \\
\hline
1 & 2.0 & 24.1 & 14.05 & 8.8 & 1010\\
1 & 2.5 & 30.0 & 11.3 & 11.5 & 1950\\
1 \& 3 & 3.0 & 36.0 & 9.45 & 14.0 & 2990\\
3 & 3.0 & 30.0 & 9.45 & 18.6 & 3000\\
\hline
\end{tabular}
\label{tab:params}
\end{table}

\section{Reimaging Optics and Detector Arrays}
\label{sec:reimage}

Over the last decade several teams have studied and implemented reimaging optics designs for large-aperture CMB telescopes and have converged on using optics tubes with multiple lenses. The upcoming instruments for the ACT \cite{henderson/etal:2015}, SPT \cite{benson/etal:2014}, and Simons Array \cite{arnold/etal:2014} will all employ an optics-tube approach in which three lenses reimage the telescope focus onto each detector array.
The detector arrays on these instruments will be superconducting bolometric detectors operated at sub-Kelvin temperatures that are sensitive to excess thermal emission. These detectors are generally optimized to achieve ``background-limited'' sensitivity, meaning that photon noise fluctuations are the dominant noise source. 

To demonstrate background-limited detector sensitivity it is important to show that the measured noise level scales as expected with changes in background loading conditions; for example, Grace et al. \cite{grace/etal:2014} show that the median detector sensitivity in the ACTPol instrument is limited by noise that scales with the atmospheric precipitable water vapor. We assume for the remainder of this work that a next generation instrument will achieve similar individual detector noise performance to ACTPol \cite{grace/etal:2014} by use of the illumination techniques described below. When CMB detectors are background-limited the best approaches for improving sensitivity are to maximize the optical efficiency, reduce the backgrounds, and increase the number of detectors while maintaining the throughput per detector. 

Coupling the detectors to a CD telescope via multiple optics tubes amplifies the CD benefits in several ways, including: 1) increasing the usable FOV diameter, 2) providing a compact cryogenic Lyot stop that mitigates spillover, simplifies baffling, and increases sensitivity by reducing the background optical load, 3) dividing a large cryostat window into smaller ones, which reduces window and lens size, thickness, and coating complexity, and 4) maximizing the number of detectors in each silicon wafer to reduce costs and increase sensitivity. The cryogenic Lyot stop also provides uniform illumination for half-wave-plate polarization modulators to mitigate systematic effects.

The detector arrays illuminated by these optics can generally be described as ``feed-coupled'' detectors with single-moded approximately Gaussian beams. The ``feed'' may be a feedhorn, a lenslet, or a phased-antenna array \cite[e.g.,][]{hanany/niemack/page:2013}. The Gaussian optics are typically designed such that a substantial fraction of the beam (10 -- 50\%) illuminates the Lyot stop, baffles, and walls surrounding the detector array. This highlights the importance of reducing the temperature of these baffles, which even at 4\,K could contribute more optical loading and photon noise than the 2.7\,K CMB. In \cite{griffin/bock/gear:2002} the tradeoffs between detector aperture and background loading levels are explored for a variety of different configurations. For  close-packed ``feed-coupled'' arrays after minimizing sources of background loading, the optimal detector aperture for measuring an extended source like the CMB is generally between 1 -- 2\,$f\lambda$, where $\lambda$ is the wavelength, which is the regime of most current detector arrays. For FOV-limited designs with no limit on the detector packing density, the optimal spacing when taking into account instrument backgrounds is typically near 1.3 -- 1.5\,$f\lambda$, while readout-limited or detector-packing-limited arrays are typically designed with apertures closer to 2\,$f\lambda$.

\begin{figure}[t]
\centering
\fbox{\includegraphics[width=\linewidth]{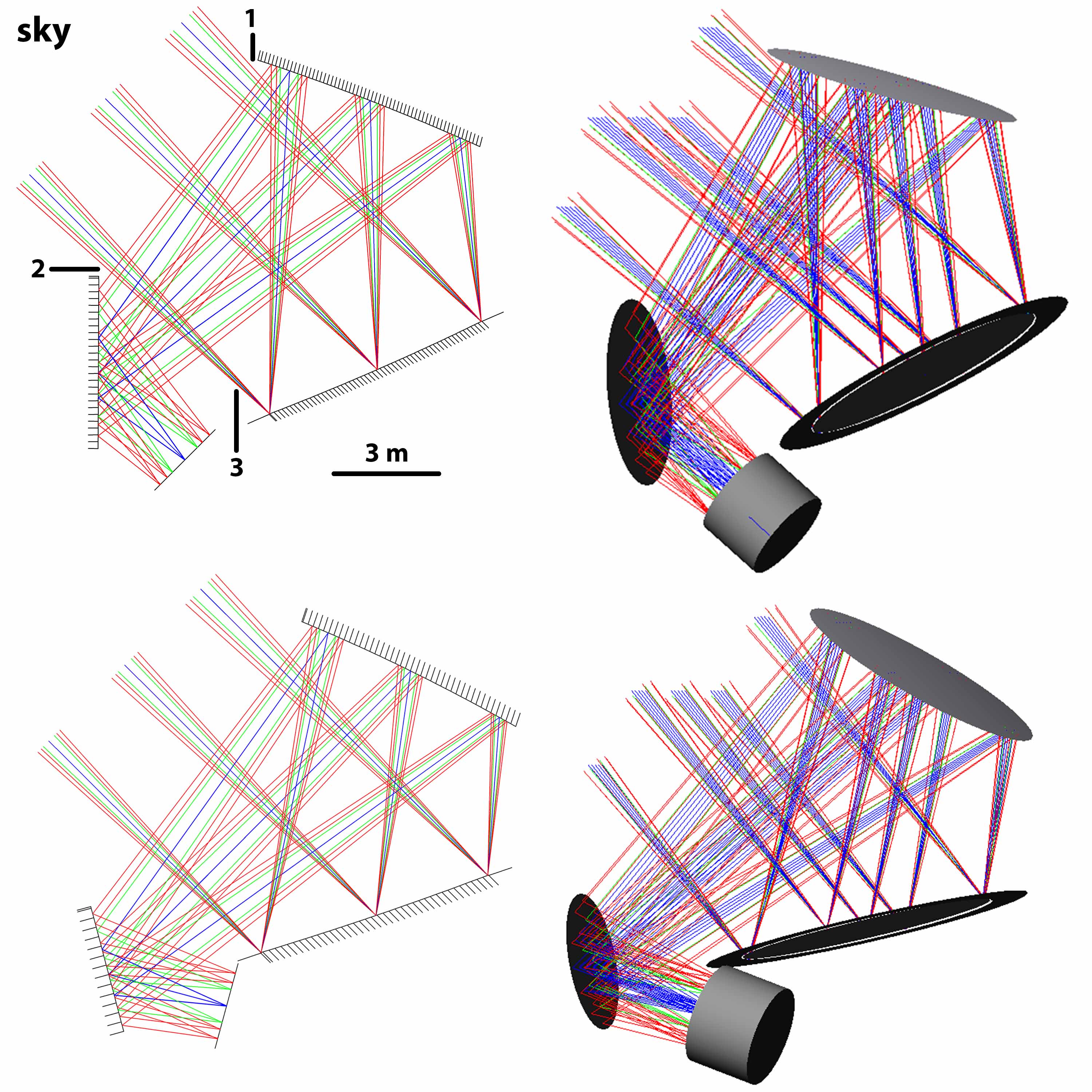}}
\caption{Two 6\,m aperture CD telescope designs with $f=3$  (top and bottom) tilted to 45$^\circ$ observing elevation. Parameters for each design are in Table~\ref{tab:params}. Both have flat tertiary mirrors (with fold angles of 45$^\circ$ for the top and 30$^\circ$ for the bottom) to move the telescope focus and receiver closer to the center of mass. The rendered images on the right also show the cylindrical receiver from Fig.~\ref{fig:receiver}. The optical clearances labeled 1, 2, and 3 can easily be adjusted by changing the telescope parameters and fold angle.}
\label{fig:folded}
\vspace{-.15in}
\end{figure}

The first multi-frequency, or multichroic, detector array was deployed in early 2015 \cite{henderson/etal:2015}, and more will be deployed soon in upcoming CMB instruments \cite{henderson/etal:2015, arnold/etal:2014, benson/etal:2014}. These arrays enable use of greater bandwidth and simultaneous observations at multiple frequency bands at the cost of building and reading out more detectors from each focal plane element. Additional challenges include that each frequency is typically coupled through the same ``feed'', which leads to tradeoffs between the optimal aperture size at each frequency. For example, \cite{datta/etal:2014} presents an analysis of optimal aperture size to maximize the mapping speed at different frequencies for the ACTPol design \cite{niemack/etal:2010}, which has $f \approx 1.35$ and a 1\,K detector cavity extending to the Lyot stop. This analysis suggests the optimal apertures for 90\,GHz, 150\,GHz, and 220\,GHz are between 1.3--1.4\,$f\lambda$, which corresponds to approximately 6\,mm, 3.7\,mm, and 2.4\,mm feed apertures, respectively.\footnote{The optimal apertures for ACTPol assume a fixed field-of-view and no constraint on the number of detectors per array.} Clearly the optimal aperture cannot be achieved for all three frequencies simultaneously. A multichroic array with two nearby frequecies can achieve the mapping speed of $\sim$1.7 optimized single frequency arrays in addition to improved spectral coverage \cite{datta/etal:2014}. This and the challenges of developing wide bandwidth optics to couple to multichroic arrays suggests a reasonable number of frequencies for a ``feed-coupled'' multichroic array is two or three.

\begin{figure}[t]
\centering
\includegraphics[width=\linewidth]{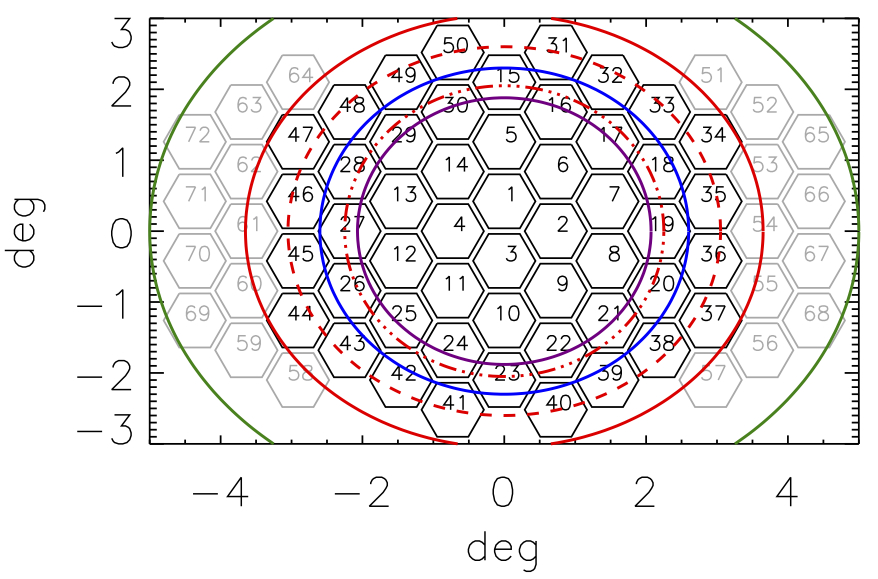}
\vspace{-.25in}
\caption{Telescope FOV for the design in Fig.~\ref{fig:folded}. The solid ellipses show the CFOV for 300\,GHz (inner/purple), 230\,GHz (blue), 150\,GHz (red), and 100\,GHz (outer/green).  The red dash and dot-dash ellipses show the FOV with 150\,GHz Strehl ratios $>0.80$ and $>0.90$, respectively.  The black hexagons indicate a possible layout for 50 optics tubes that can achieve diffraction-limited performance at 150\,GHz. The additional 22 grey hexagons on the left and right indicate possible locations for additional optics tubes that would be diffraction-limited at 100\,GHz and at lower foreground frequencies, such as 30\,GHz and 40\,GHz. Each hexagon represents a 280\,mm diameter hexagonal shaped lens with 25\,mm gaps between lenses. The plate scale is approximately 320\,mm/deg.}
\label{fig:hex}
\vspace{-.15in}
\end{figure}

The multichroic detector arrays for upcoming large-aperture CMB instruments are all expected to be fabricated on 150\,mm diameter silicon wafers.\footnote{Previously deployed detectors were fabricated on 75\,mm or 100\,mm wafers.} Superconducting detector fabrication is challenging and costly due to detector complexity and strict uniformity requirements \cite[e.g.,][]{henderson/etal:2015}; therefore, it is generally advantageous to make each wafer as sensitive as possible by maximizing the number of detectors per wafer, so long as the additional detectors do not complicate fabrication. All of the upcoming large-aperture projects plan to use hex-packed detectors on hexagonal-shaped detector wafers surrounded by wirebond pads to read out the detectors. In Advanced ACTpol the densest of these arrays will have 2012 superconducting detectors operating at 150\,GHz and 230\,GHz with 4.65\,mm apertures \cite{henderson/etal:2015}. For SPT-3G each detector wafer will have a slightly smaller number of detectors operating at 95\,GHz, 150\,GHz, and 220\,GHz, though a larger total number of detectors will be integrated into the focal plane from $\sim$10 wafers \cite{benson/etal:2014}. These designs are approaching the practical limits of both wirebond density and detector-packing density, though a factor of $\sim$1.5 increase in detector count per 150\,mm wafer may be achieved for a Stage IV experiment. For the purpose of this study, we assume that next generation detectors will have 2000 -- 3000 detectors per 150\,mm wafer.

These considerations guide the design of the reimaging optics. Around the perimeter of each hexagonal detector wafer there is inevitably some dead space (e.g. bond pads and mechanical structure) that does not couple light to detectors. This dead space decreases the effective system throughput. By designing hex-packed optics tubes that each couple to a single detector wafer, we remove the dead space from around the detector arrays and provide more space for support structures, while introducing a small amount of dead space into the telescope focal plane. To minimize the dead space between optics tubes, the first modular element in each tube is a hexagonal refractive lens that captures and collimates the telescope beam before it diverges beyond the telescope focus. Based on the design in Fig.~\ref{fig:folded}, we divide the CFOV into 50 hexagonal fields, each with a maximum dimension of 280\,mm and a 25\,mm space for structural support between neighboring lenses (Fig.~\ref{fig:hex}). This geometry provides roughly 90\% active area with greater throughput than the 150\,GHz DLFOV, which is $>10\times$ the ACTPol throughput (Fig.~\ref{fig:hex}).

\begin{figure}[t]
\centering
\fbox{\includegraphics[width=\linewidth]{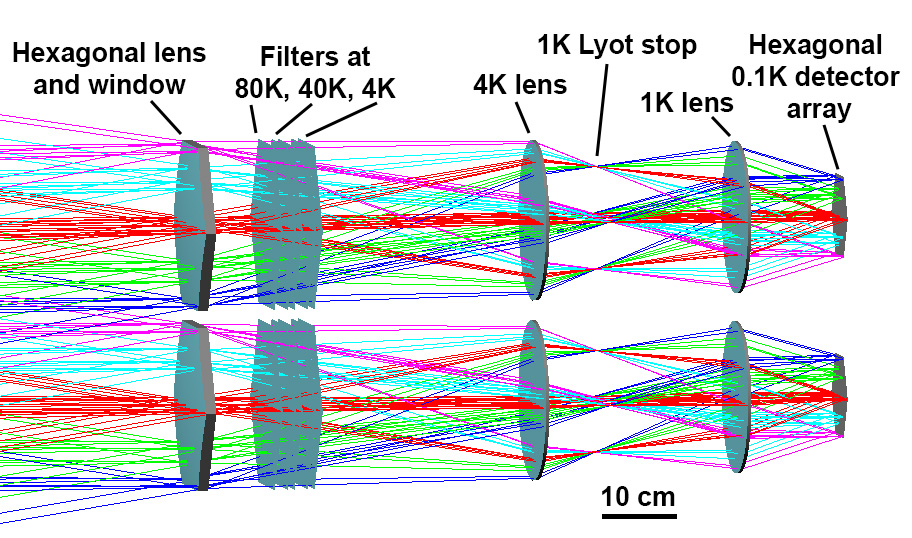}}
\caption{Two neighboring sets of reimaging optics that speed up the telescope focus from $f = 3$ to $f = 1.5$ and provide a Lyot stop to control primary mirror illumination and spillover. The hexagonal silicon lens is at the telescope focus and is also used as a room-temperature vacuum window. The two cryogenic lenses are circular and fit inside the footprint of the hexagonal lens to facilitate thermal isolation between temperature stages.}
\label{fig:tube}
\vspace{-.15in}
\end{figure}

\begin{figure}[t]
\centering
\includegraphics[width=\linewidth]{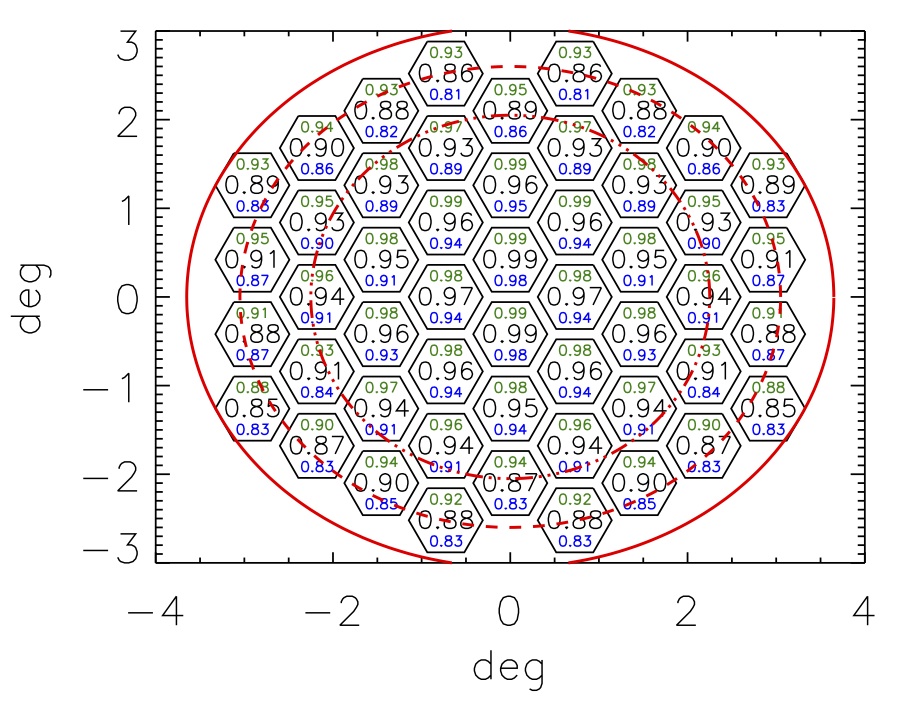}
\caption{Results of Strehl ratio calculations at 150\,GHz at the detector array focus for each of the 50 central optics tubes in Fig.~\ref{fig:hex}. Strehl ratios are calculated for the 15 fields ray traced in Fig.~\ref{fig:tube} in each optics tube. The largest (black) value in each hexagon is the average for all 15 fields, while the upper (green) value is the maximum, and the lower (blue) value is the minimum. The CD ellipses for 150\,GHz Strehl ratios of $0.7$, $0.8$, and $0.9$ from Fig.~\ref{fig:hex} are shown in red, clearly demonstrating the image quality improvement from the reimaging optics.}
\label{fig:strehls}
\vspace{-.15in}
\end{figure}

After the hexagonal lens, each subsequent optical aperture and cryogenic support structure must fit inside the footprint of the first lens to prevent interference between optics tubes, as shown in Fig.~\ref{fig:tube} (and Fig.~\ref{fig:receiver}). One cryogenic lens helps to optimize the Lyot stop and image quality, and a final lens improves the image quality and telecentricity, while focusing at $f = 1.5$ onto the 140\,mm wide active area of a hexagonal detector array. The second and third lenses are not at a focus, so they have circular instead of hexagonal apertures, and therefore must be smaller diameter than the smallest dimension of the hexagonal lens (242\,mm). The design shown in Fig.~\ref{fig:tube} meets these requirements and uses silicon lenses \cite{datta/etal:2013}\footnote{Similar optical performance is likely achievable with alumina lenses.} to improve on the local image quality at the telescope focus. As shown in Fig.~\ref{fig:strehls}, nearly identical optics tube designs achieve minimum Strehl ratios $>0.80$ and average Strehl ratios $>0.85$ at 150\,GHz for all of optics tubes 1 -- 50 in Fig.~\ref{fig:hex}. 

A different receiver optics approach based on larger-diameter (720\,mm) alumina lenses is being pursued for the SPT-3G instrument \cite{benson/etal:2014}. This approach is appealing from the optics design perspective and appropriate for current instruments; however, it presents challenges for next generation instruments that can be avoided by use of close-packed optics tubes. Larger-diameter lenses are necessarily thicker at the center, which results in greater loss within each refractive element. Having a single optics tube also requires development of wider-bandwidth high-efficiency coatings for all optical elements. In comparison, the optics tubes in Fig.~\ref{fig:hex} can easily be divided for use over a wide frequency range; for example, tubes 1 -- 14 could be used with multichroic 220/300\,GHz arrays (or lower frequencies), tubes 15 -- 50 with 100/150\,GHz arrays, and tubes 51 -- 72 with 30/40\,GHz (or up to 100\,GHz) arrays. 

With the close-packed optics tube approach described here, if detector array designs for upcoming instruments with $\sim$2000 detectors per wafer were installed into optics tubes 1 -- 50 in Fig.~\ref{fig:hex}, that would provide roughly $10^5$ detectors on one telescope. However, with $f = 1.5$ at the optics tube focus, the mapping speed could be increased further by modestly increasing the number of detectors on each wafer. For example, if either a third frequency were added to the Advanced ACTPol detector arrays or the feed aperture for the multichroic 150/230\,GHz detectors was decreased from 4.65\,mm to 3.9\,mm (resulting in $\sim1.3$\,$f\lambda$ at 150\,GHz and $\sim2$\,$f\lambda$ at 230\,GHz), it could provide roughly 3000 detectors per wafer with a good balance of aperture efficiencies. In this case, or by including optics tubes 51 -- 72 in Fig.~\ref{fig:hex} at lower frequencies, the optics design presented here could illuminate $>$150,000 diffraction-limited superconducting detectors.

\section{Systematic Considerations}
\label{sec:systematics}

Off-axis Gregorian telescopes are widely used in current CMB research \cite[e.g.][]{henderson/etal:2015,benson/etal:2014, arnold/etal:2014}. While the CD optics designs presented here explore a new regime in CMB optics, in terms of systematic considerations they are relatively straightforward extensions of approaches used in current CMB telescopes. Several studies of the systematic requirements for CMB polarization measurements have been done \cite[e.g.,][]{hu/hedman/zaldarriaga:2003, odea/challinor/johnson:2007, miller/shimon/keating:2009}, and recent measurement results demonstrate that instruments with similar systematics to these designs can measure B-mode polarization. 

Polarization systematics can broadly be categorized as polarization distortions or leakage of temperature into polarization. Hu et al. \cite{hu/hedman/zaldarriaga:2003} suggest that polarization distortion systematics must be controlled at the $10^{-2}$ -- $10^{-3}$ level and leakage of temperature into polarization at the $10^{-3}$ -- $10^{-4}$ level to achieve the minimum detectable level of inflationary B-mode polarization at degree angular scales. These are useful benchmarks for instrument design, though it is worth noting that modulation and calibration techniques can be used to mitigate instrumental distortions exceeding these levels. For example, using rotating half-wave plate polarization modulators, building a telescope capable of boresight rotations, and scanning the sky at multiple parallactic angles are three modulation approaches for calibrating and removing polarization systematics. In addition, Yadav et al. \cite{yadav/su/zaldarriaga:2010} present a method for removing systematic instrumental distortions like these from the data by using unbiased estimators to extract the off-diagonal correlations in the CMB polarization. 

We focus on potential sources of systematic contamination from the CD telescope optics, because lenses and detector ``feed'' elements have been used in every measurement of CMB B-mode polarization thus far \cite[e.g.][]{hanson/etal:2013, bkp:2015, polarbear:2014, vanengelen/etal:2015} and are not unique to the designs presented here. We quantify the telescope differential gain between x and y polarizations, which can leak temperature into polarization signals, as well as the cross polarization response (crosspol), which can distort polarization measurements. Both are calculated using $GRASP$ physical optics software for the primary and secondary mirrors of the upper telescope in Fig.~\ref{fig:folded} at 150 GHz. Calculations are performed with detectors at the center and edges ($\pm 3.1 ^\circ$) of the CD focal plane and for feedhorn edge tapers of $-12$\,dB and $-3$\,dB. The differential gain calculations include the finite conductivity of the aluminum mirrors, $2.5\times10^7$\,S/m \cite{tran/etal:2008}, which is known to cause differential gain between x and y polarizations. The differential gain was found to be less than $10^{-4}$ at multiple field points spanning the CFOV. The tertiary fold mirror is expected to increase the differential gain by an amount that scales with the angle of incidence. The average tertiary angle of incidence for the lower design in Fig.~\ref{fig:folded} is smaller than the angle of incidence on the primary mirror and therefore the fold mirror in this design is expected to introduce less differential gain than the primary mirror.
Crosspol calculations are performed for x and y polarizations and the worse response is reported. The results are summarized in Table~\ref{tab:crosspol}.  
Crosspol is below $-80$\,dB at the center of the field and rises to $-45$\,dB near the edge of the CFOV. The worst crosspol in this telescope design is much better than the crosspol of most feedhorns and substantially below the target in Hu et al. \cite{hu/hedman/zaldarriaga:2003}. In summary, crosspol and differential gain from the telescope are well within the acceptable limits for CMB polarization measurements. 

For comparison, Tran et al. \cite{tran/etal:2008} present crosspol and differential gain analysis for a CD design with $f \approx 1.5$. They similarly find that the CD telescope has $-45$\,dB crosspol near the perimeter of the CFOV, where the Strehl ratios drop to 0.7. All the analyses presented in \cite{tran/etal:2008} show that the CD design meets the requirements for CMB polarization measurements and outperforms a Gregorian design with similar $f$ and aperture. Unlike traditional Gregorian telescopes, these CD designs may also be sufficiently compact to mitigate instrumental systematics by rotating the boresight of the entire telescope around the optical axis.

\begin{table}[b]
\vspace{-.3in}
\centering
\caption{Physical optics analysis for upper telescope in Fig.~\ref{fig:folded}.  }
\begin{tabular}{cccc}
\hline
Azimuth & Elevation  & Taper  & Crosspol  \\
angle ($^\circ$) & angle ($^\circ$) & (dB) & (dB)  \\
\hline
0 & 0 & -12 & -88 \\
0 & $\pm$3.1 & -12 & -56 \\
$\pm$3.1 & 0 & -12 & -45\\
0 & 0 & -3 & -84 \\
0 & $\pm$3.1 & -3 & -55  \\
$\pm$3.1 & 0 & -3 & -45  \\
\hline
\end{tabular}
\label{tab:crosspol}
\end{table}

\section{Engineering Considerations}
\label{sec:eng}

To achieve background-limited noise performance, bolometric CMB detectors must be cooled to sub-Kelvin temperatures. This makes the relationship between the optics and cryogenic receiver design one of the primary engineering challenges for CMB instruments. 

Silicon lenses are placed at the telescope focus to minimize FOV loss in this optics design. To take full advantage of this, a large vacuum window must cover all the lenses or the lenses must act as vacuum windows themselves. With a perimeter thickness of roughly 10\,mm, the 280\,mm diameter crystalline silicon lenses are sufficiently strong to support atmospheric pressure and are used as vacuum windows in this design. The loss tangent of silicon at millimeter wavelengths varies with doping and with temperature. The silicon lenses used in ACTPol were shown to have a loss tangent of $\tan \delta \approx 10^{-4}$ at room temperature and smaller below 10\,K \cite{datta/etal:2013}. Interestingly Krupka et al. \cite{krupka/etal:2006} show that some types of silicon have a $\tan \delta$ minimum near 300\,K and a lower minimum below 20\,K, which supports the use of silicon lenses as vacuum windows instead of operating them at an intermediate temperature between roughly 40\,K -- 200\,K. The silicon loss tangent can temporarily degrade if ultraviolet radiation excites charge carriers into the conduction band, but we have shown that simply shading the silicon from direct sunlight or covering it with a 10\,mm thick sheet of Zotefoam sufficiently mitigates this effect.\footnote{Silicon conductivity measurements were done at the ACT site by B. Koopman.}

Behind the vacuum windows thermal-blocking filters and low-pass filters are used to prevent infrared and submillimeter radiation from entering the receiver and heating the colder stages. Fig.~\ref{fig:tube} shows the optical filters to reject this higher frequency radiation as well as the temperatures of each optical element. We estimate roughly 200\,W of radiative power will be absorbed by the 80\,K stage. Due to the larger window area this is a substantially larger radiative load than current instruments, but it can easily be intercepted by adding single-stage pulse tube refrigerators designed to operate at 80\,K. Additional thermal-blocking and low-pass filters are installed on the subsequent 40\,K and 4\,K temperature stages, which are cooled using two-stage pulse tube refrigerators like those operating on current instruments. The intermediate 80\,K stage dramatically reduces the load on the 40\,K stage compared to current instruments, enabling the two stage pulse tube operation to be optimized for maximum cooling power at 4\,K. 

\begin{figure}[t]
\centering
\fbox{\includegraphics[width=\linewidth]{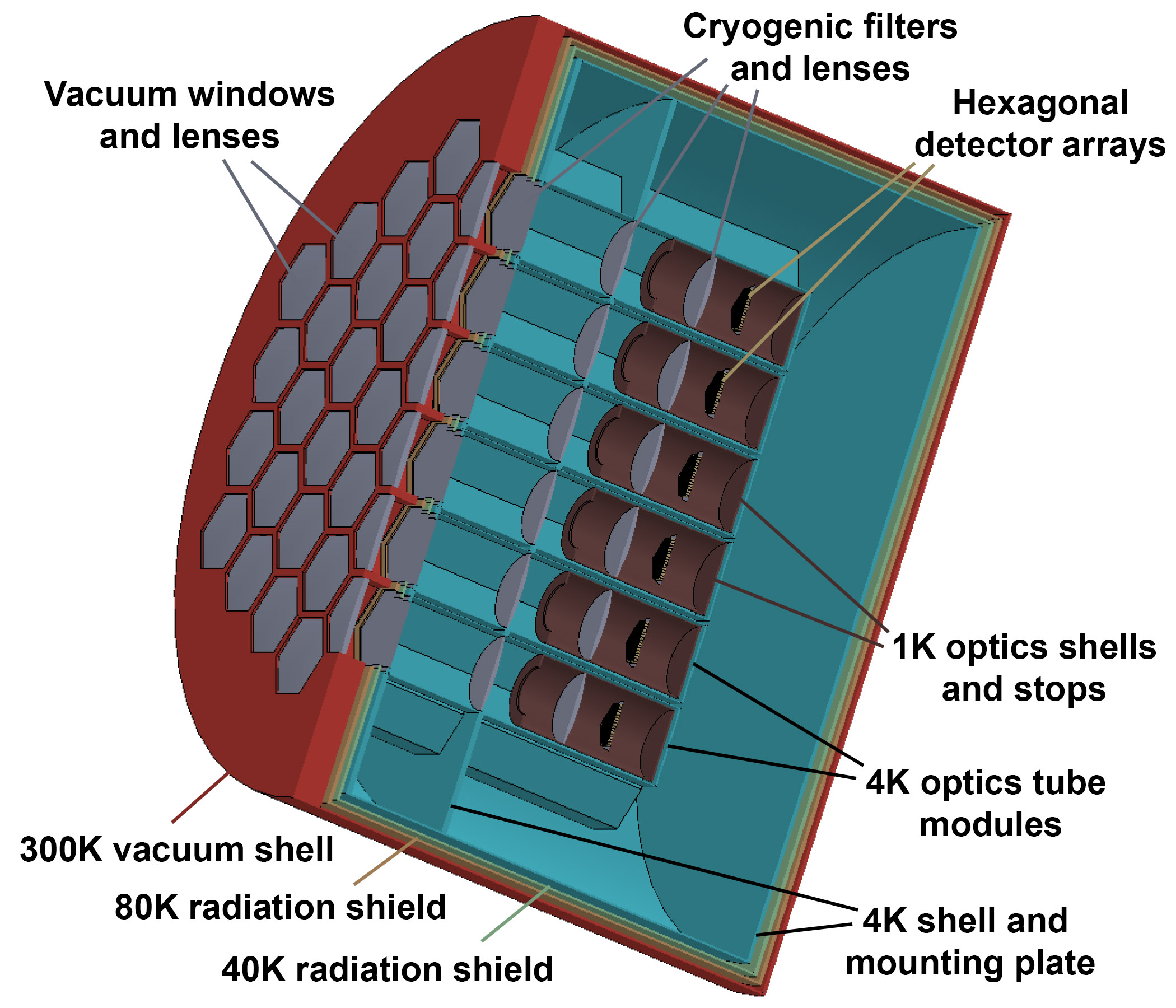}}
\caption{Cross-section of a preliminary receiver design for 50 optics tubes and detector arrays. The vacuum cylinder is 2.5\,m diameter by 1.8\,m long as shown in Fig.~\ref{fig:folded}. A honeycomb structure is used to support the room temperature lenses that also serve as vacuum windows. 80\,K and 40\,K radiation shields surround the lower temperature optics tubes and support hexagonal infrared blocking filters to reduce the radiation load on the 4\,K stage. Each of the optics tube modules attaches to the 4\,K mounting plate, enabling individual tubes to be deployed independently. Inside each optics tube is a 1\,K optics shell that cools the Lyot stop and the final lens to minimize background loading on the 0.1\,K hexagonal detector arrays. {\it (Receiver design courtesy of S. W. Henderson.)}}
\label{fig:receiver}
\vspace{-.15in}
\end{figure}

Fig.~\ref{fig:receiver} shows a preliminary receiver design with cylindrical shells at both 40\,K and 4\,K for mounting additional filters and to minimize radiative load. The remaining optical components and detector arrays are assembled into modular optics tubes that are mounted onto a 4\,K mounting plate. The final stages of cooling for the optics and detector arrays are provided by a pulse-tube-cooled dilution refrigerator (DR). This enabling technology operates continuously and reaches lower temperatures with many times more cooling power (e.g. 400\,$\mu$W of cooling power at 100\,mK) than helium-3 sorption refrigerators and adiabatic demagnetization refrigerators.

The modularity of the optics tubes is a substantial engineering advantage of this approach. After the receiver shells are assembled, individual optics tubes can be added, repaired, or replaced as needed without the risk of damaging other optics. In addition, each optics tube could operate at a different set of observing frequencies. Each tube consists of several modular components. A 4\,K shell is used to mount it onto the receiver and to support the 4\,K lens. A 1\,K shell  defines the Lyot stop to control illumination of the primary mirror, terminates spillover from the detector array, and supports the 1\,K lens. The 4\,K and 1\,K shells can be designed with reentrant supports \cite[e.g.,][]{thornton/etal:2008} to provide sufficient thermal isolation. Inside the 1\,K shell there will be a thermal isolation support structure for the 0.1\,K hexagonal detector array. 

The 50 optics tubes analyzed in Fig.~\ref{fig:strehls} and depicted in Fig.~\ref{fig:receiver} are mechanically identical. The three silicon lenses in each tube are plano-convex and aspheric on the convex side. To achieve the image quality presented in Fig.~\ref{fig:strehls}, five different combinations of the aspheric surfaces for the two cryogenic lenses were optimized, leading to a total of eleven different lens designs. In other words, the lenses that act as vacuum windows are all identical, and there are five pairs cryogenic lens designs, while all 50 optics tubes have identical mechanical structures. Fig.~\ref{fig:receiver} shows the critical optical components of the receiver, including a hexagonal feedhorn array design \cite{henderson/etal:2015}. The receiver layout is similar in concept to the ACTPol cryostat \cite{niemack/etal:2010} in which the optics tubes are modular and there is space behind the optics tubes for the pulse tubes and dilution refrigerator. The thermal connections to cool the detector arrays and 1\,K components will be made through holes in the rear of the optics tubes. A preliminary assessment of the tolerance requirements for the telescope, reimaging optics, and detectors indicates similar requirements to the ACT \cite{fowler/etal:2007}.

The design presented here demonstrates the feasibility of building a telescope and receiver to map the CMB $10\times$ faster than upcoming observatories. More detailed collaborative systematics and engineering studies are needed before the selection of a next generation telescope or receiver design is finalized for a Stage IV CMB observatory.

\section{Conclusions}
\label{sec:conclusion}

We have shown that one large-aperture CMB telescope can illuminate $>10\times$ more detectors than upcoming CMB instruments \cite{arnold/etal:2014, benson/etal:2014, henderson/etal:2015} and provide approximately $10\times$ faster mapping speed. Our approach is based on a crossed-Dragone telescope design with modular close-packed optics tubes filling the telescope focal plane to illuminate $> 10^5$ detectors. This demonstrates for the first time that a Stage IV CMB observatory with $10^5$ -- $10^6$ detectors can be built using a small number of telescopes.

\vspace{-.05in}
\section*{Funding Information:}
National Science Foundation Directorate for Mathematical and Physical Sciences (1454881) and Cornell University.
\vspace{-.05in}

\section*{Acknowledgments}
The author thanks S. W. Henderson for creating the receiver model (Fig.~\ref{fig:receiver}) as well as R. Dunner and P. Rojas for use and support of $GRASP$ for physical optics calculations. Several people provided useful discussions and/or comments, including L. A. Page,  A. T. Lee, S. T. Staggs, S. Dicker, G. Stacey, N. W. Halverson, J. J. McMahon, M. J. Devlin, S. Padin, J. E. Carlstrom, J. W. Fowler, S. Hanany, S. Parshley, S. W. Henderson, B. J. Koopman, S. M. Bruno, and attendees of the 2015 CMB-S4 meeting. The author also appreciates the helpful feedback from the reviewers. 
\vspace{-.05in}

\bibliography{full_bib}

\end{document}